\documentclass[12pt,a4paper,english]{article}
\usepackage[T1]{fontenc}
\usepackage[latin9]{inputenc}



\usepackage{babel}

\usepackage{cites,setspace,url}
\usepackage{epsfig}

\usepackage{rotating}

\bibnum{k}
\begin{document}

\title{Note on Drexel tests of the IMB R1408 PMTs used in the inner veto of both far and near detectors of the Double Chooz experiment}

\author{Karim ZBIRI}

\date{Drexel University}
\maketitle

\begin{abstract}
This report is related to the testing performed at Drexel University of 250
R1408 PMTs. Based on this testing, 156 R1408 PMTs were chosen to be used in
the inner veto of the detectors of the Double Chooz experiment\cite{doublechooz}.

\pagenumbering{roman}
\tableofcontents{}
\setcounter{page}{1}
\end{abstract}

\pagenumbering{arabic}
\setcounter{page}{1}


\newpage


\section{Introduction}

More than 260 R1408 PMTs were obtained from UC Irvine, these PMTs were used in
IMB\cite{IMB} and SuperKamoika\cite{superkamioka} experiments.
These 13 stages, 8 inches, R1408 PMTs don't have a single photo-electron peak and the cause of that
is expected to be due to their venetian blind dynodes.
The aim of the testing performed at Drexel University was to be sure that
most of these PMTs are still working correctly, and then make a choice of
156 of them. The testing concerned the determination of the high
voltage corresponding to a gain of $\sim10^{7}$,
saturation level and the dark noise rate.
To make this testing different calibration techniques were used, the
photo-statistics and the light pulser methods. Most of these
PMTs came with their original linear bases, a first testing of 250 PMTs
was made with these bases, after that a new
testing was performed with a new tapered bases.

\section{Calibration techniques}
\subsection{The Photo-statistics method}

\begin{figure}[h]
\begin{center}\includegraphics[%
  scale=0.7]{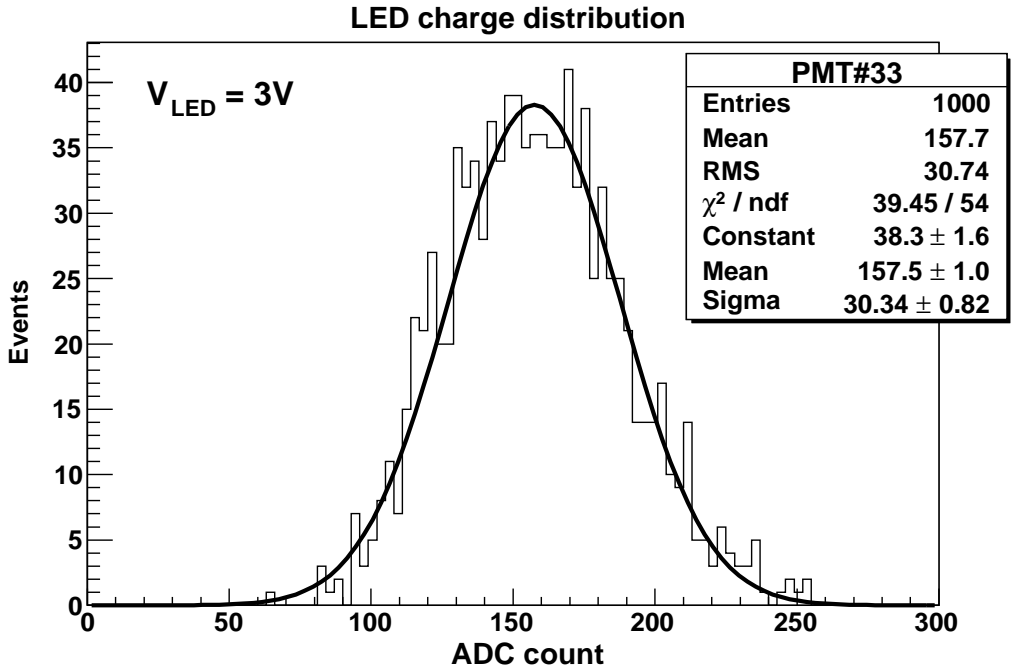}\end{center}

\caption{\label{cap:LED-pulse-pmt33} ADC distribution for LED pulses.}
\end{figure}

\begin{figure}[h]
\begin{center}\includegraphics[%
  scale=0.7]{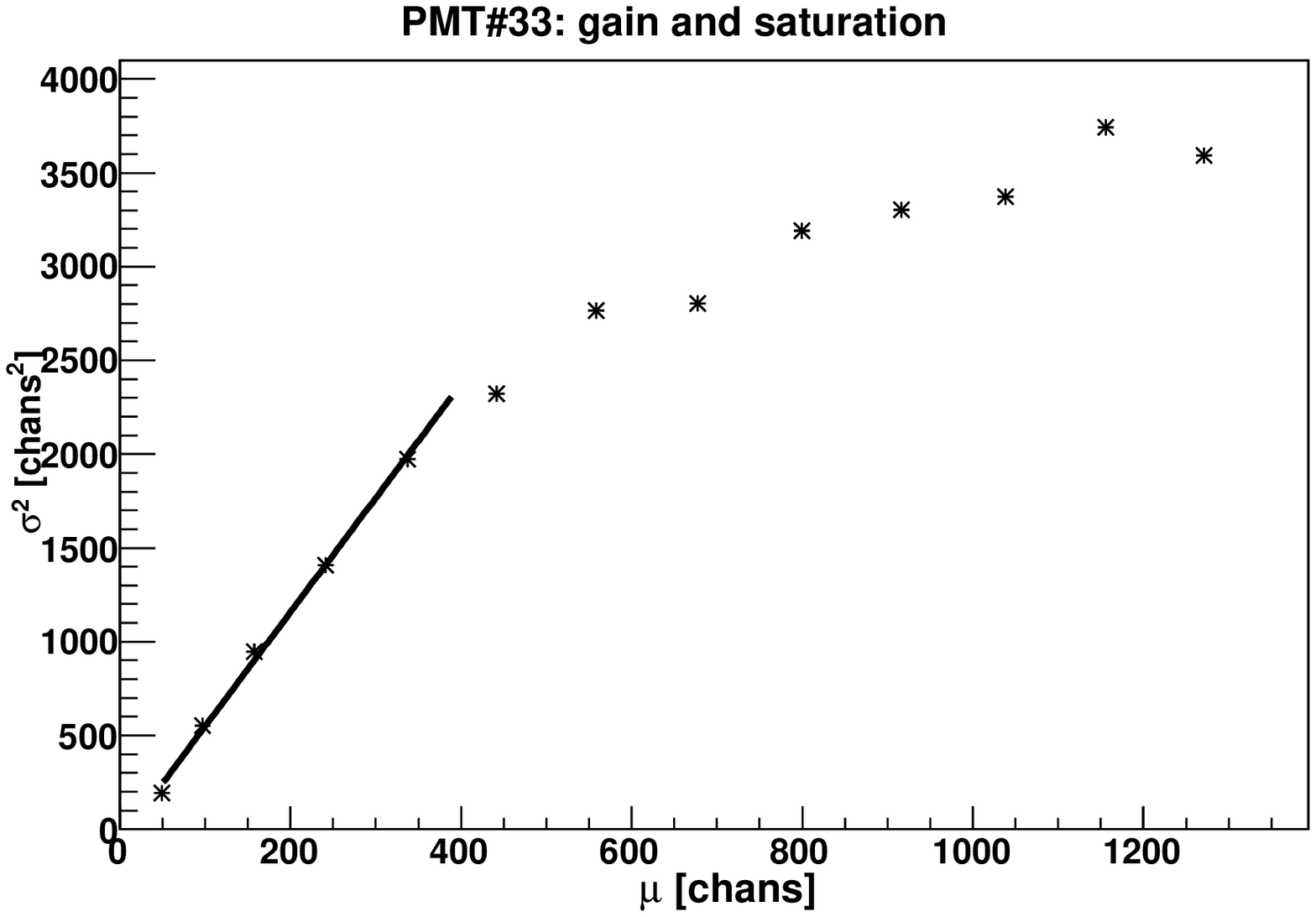}\end{center}

\caption{\label{cap:PMT33-gain} ADC variance($\sigma ^2$) versus ADC
mean($\mu$) for a variety pulse heights. The slope of the fit of the linear
region gives the gain, and the
$saturation~level=441/3.04=145~photo-electrons$.}
\end{figure}

 In this section I will summarize the principle of the Photo-statistics method.\\
With an ADC we are not measuring a charge q directly, but an ADC value a that
is related to q by:
\begin{equation}
\label{ADCGain}
a=q+a_{p}
\end{equation}
Where $a_{p}$ is the ADC pedestal.
As it can be demonstrated, the gain can be extracted via
the following relation:
\begin{equation}
\label{gain}
\frac{d \sigma _a^2}{d \mu _a}=2Ge
\end{equation}
Where $\mu _a$ and $\sigma _a$ are the mean and the variance of ADC charge
distribution, and Ge is the ADC count per photo-electron, called also
gain/channel, with $e=1.6~10^{-19}C$. All we need is
to take a series of LED/laser runs at varying light levels, and for each of
these runs to measure the statistical mean and variance of spectra as shown
in figure \ref{cap:LED-pulse-pmt33}, after that plot the variance versus
the mean, and fit the linear region with a straight line like
in figure \ref{cap:PMT33-gain}. The gain will be extracted from
the slope of the line following the equation \ref{gain}.
One important feature of this method is it is
electronics pedestal independent, to get the absolute gain one needs
to multiply the gain/channel by the ADC coefficient calibration.
To obtain the saturation level I
consider the point after the linearity of PMT response is broken, the ratio
of the ADC mean at the considered point and the gain/channel gives
the saturation
level in the number of photo-electrons, figure \ref{cap:PMT33-gain} shows
an example of this procedure. Since the saturation level is determined by
eyes it becomes human dependent, two measurements performed by two different
persons may give different values.\\

\subsection{The light pulser method}
\subsubsection{Characteristics of the light pulser}
An ${}^{241}\hbox{\rm Am}$ ${\alpha}$ source is fixed on a NaI crystal.
The choice of the crystal is very important in the way that it will let the
emitted ${\alpha}$ particles to have a uniform energy loss which allows the
production of monochromatic photons inside the scintillator. The scintillator
crystal chosen for the present measurements is NaI(Tl), The ${\alpha}$ peak
appears at an equivalent ${\gamma}$ ray energy of 2.6 MeV. The table
\ref{cap:alpha source} summarizes the characteristics of the used Am pulser.\\
\begin{table}[h]
\begin{center}\begin{tabular}{|c|c|}
\hline
Isotope&
${}^{241}\hbox{\rm Am}$ \tabularnewline
\hline
Half-life&
458 years\tabularnewline
\hline
Emission&
${\alpha}$ particles, 59.5 keV Np L-X-rays\tabularnewline
\hline
Gamma Equivalent Energy (G.E.E)&
1.5 to 3.5 MeV\tabularnewline
\hline
Tolerance on G.E.E&
${\pm}15{\%}$\tabularnewline
\hline
Count Rate&
10 to 2000 cps\tabularnewline
\hline
FWHM:&
$3$ to $5{\%}$\tabularnewline
\hline
Temperature Range&
$4^{\circ}C$ to $40^{\circ}C$\tabularnewline
\hline
\end{tabular}\end{center}
\caption{\label{cap:alpha source} Properties of the Am pulser}
\end{table}

\begin{figure}[h]
\begin{center}\includegraphics[%
  scale=0.5]{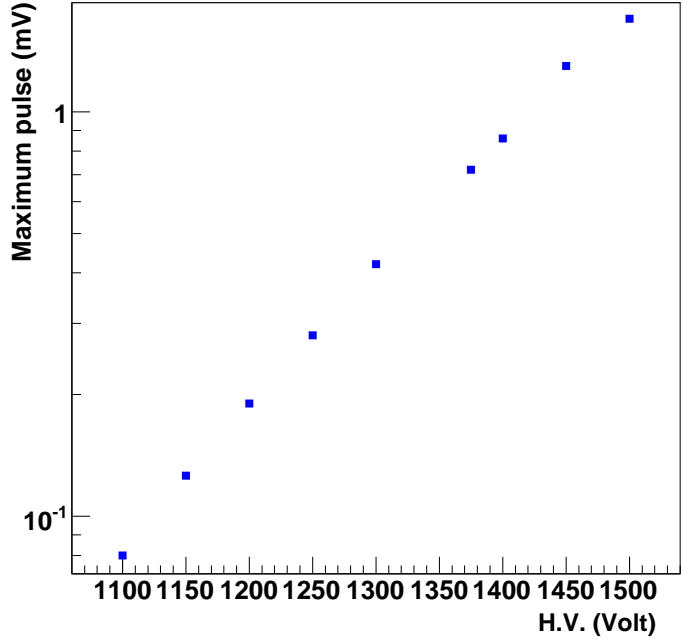}\end{center}

\caption{\label{cap:maxpulse-HV} Evolution of the maximum pulse with the
applied high voltage for the PMT reference (using the original linear base).}
\end{figure}

\begin{figure}
\begin{center}\includegraphics[%
  scale=0.5]{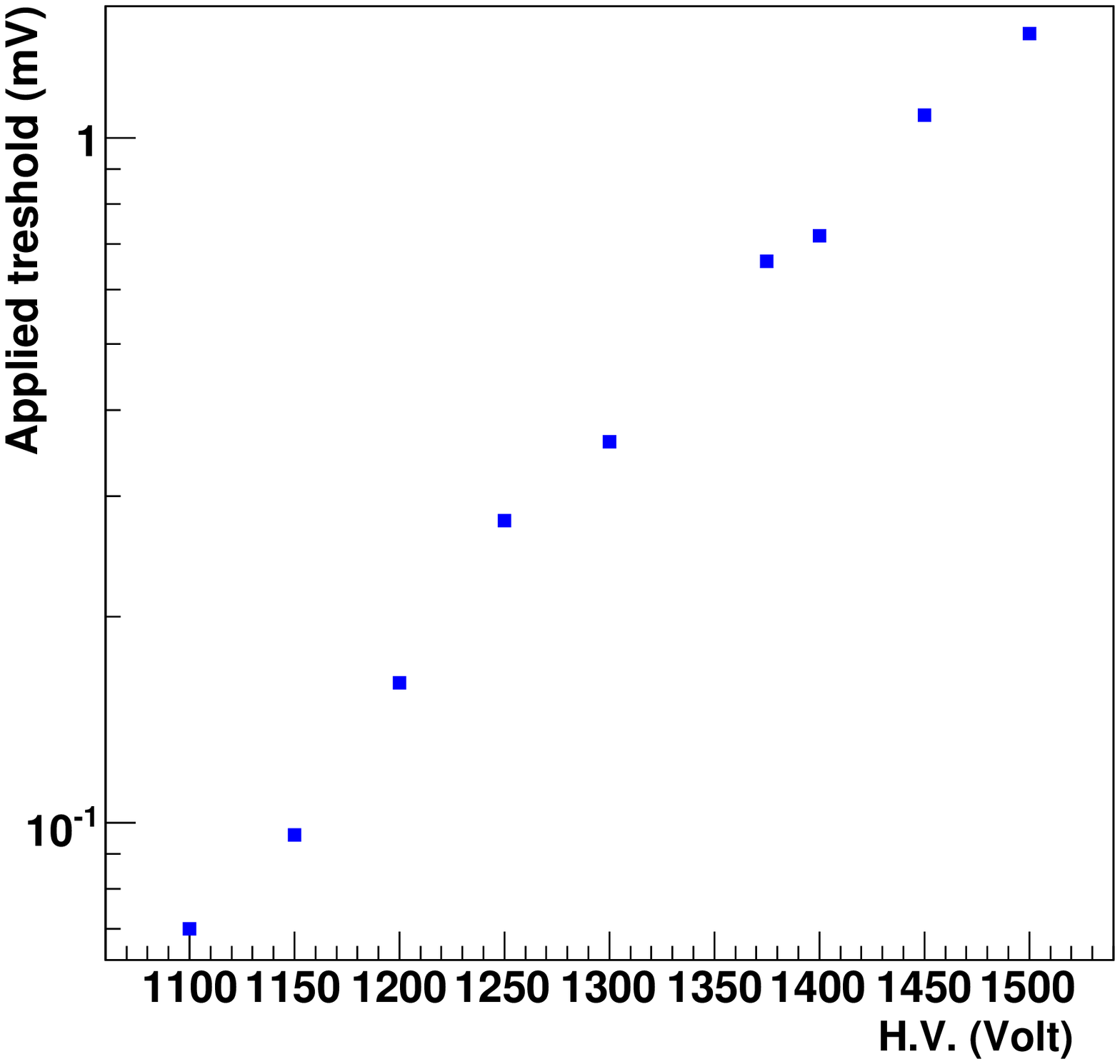}\end{center}

\caption{\label{cap:threshold-HV} Evolution of the Threshold value with the
applied high voltage for the PMT reference(using the original linear base).}
\end{figure}



\subsubsection{The Procedure of the gain determination}

I started by testing with the photo-statistics method a set of 8 PMTs from
which I selected the most convenient one from the point of view of low dark
noise rate and high saturation level. The chosen PMT became the PMT reference
for the measurements with the Am pulser. The figure \ref{cap:maxpulse-HV} shows
the evolution of the maximum pulse of the PMT reference with the applied high
voltage. The maximum pulse increase in exponential way with increasing applied
high voltage. One important parameter is the value of the trigger which need to
be set properly for accurate determination of the maximum pulse. The figure
\ref{cap:threshold-HV} shows the variation of the threshold value with
variable high voltage. To obtain the maximum pulse I used a scope tuned to its
maximum averaging value. \\
The photo-statistics method gave the high voltage
corresponding to the gain $\sim10^{7}$ for the PMT reference at 1375 Volts,
the maximum pulse corresponding to this high voltage is 720 mV as we can see
from the figure \ref{cap:maxpulse-HV}. To get the high voltage corresponding
to the gain $\sim10^{7}$ for the other PMTs, I just increased the value of
the  high voltage applied to each one until I obtained the maximum pulse
of $\approx 720 \, \hbox{\rm mV}$, then I considered these PMTs tuned to the appropriate gain
value. \\
To keep the light pulser at fixed distance from the photocathod,
the ${}^{241}\hbox{\rm Am}$ source was fixed on the top of a funnel.


\section{PMTs testing results}
\begin{figure}
\begin{center}\includegraphics[%
 scale=0.5 ]{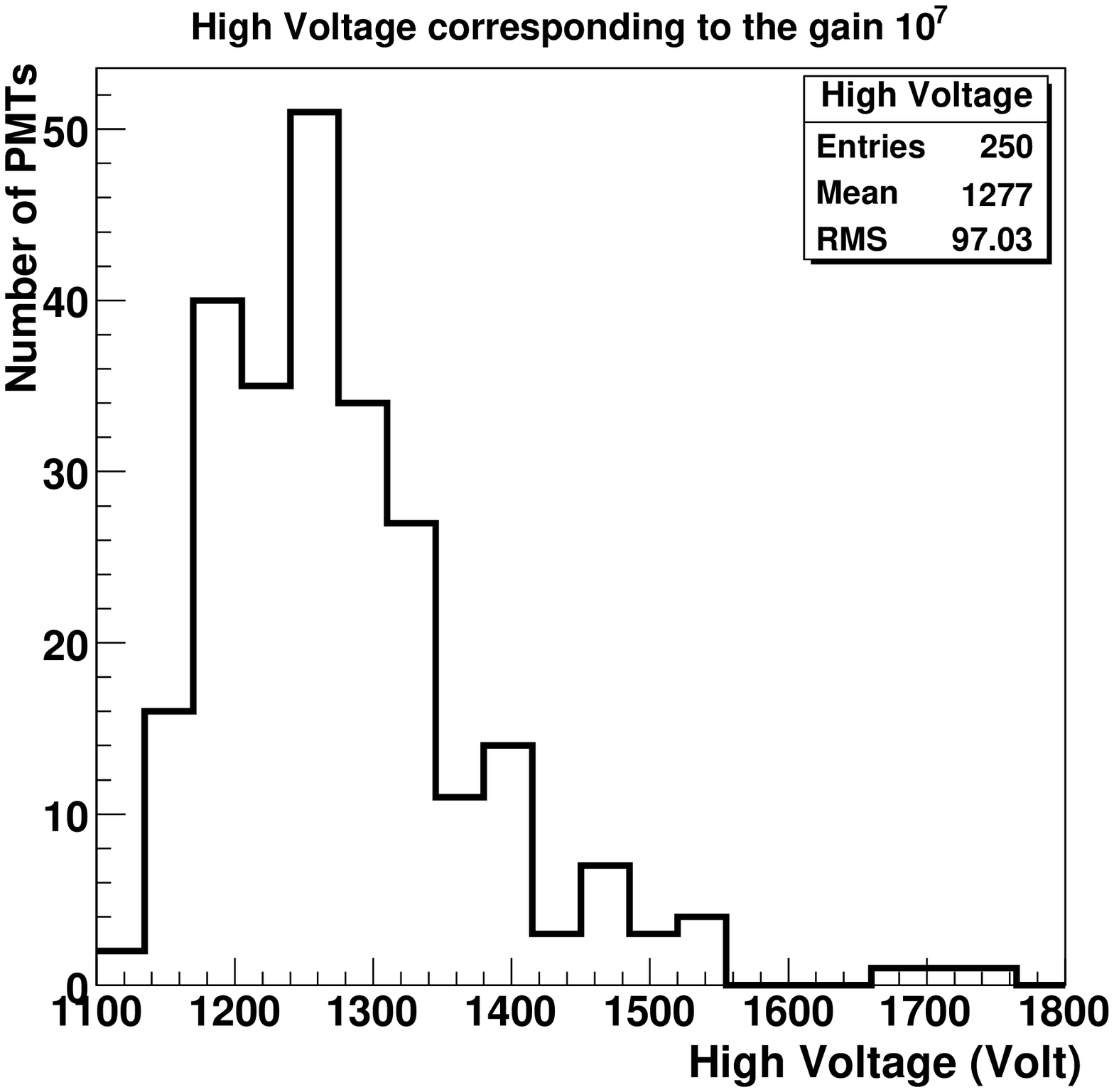}\end{center}
\caption{\label{cap:HV_OLDBase} The high voltage corresponding to the gain
$\sim10^7$ obtained with the light pulser(using the original linear bases).}
\end{figure}

\begin{figure}
\begin{center}\includegraphics[%
 angle=-1,scale=1.2 ]{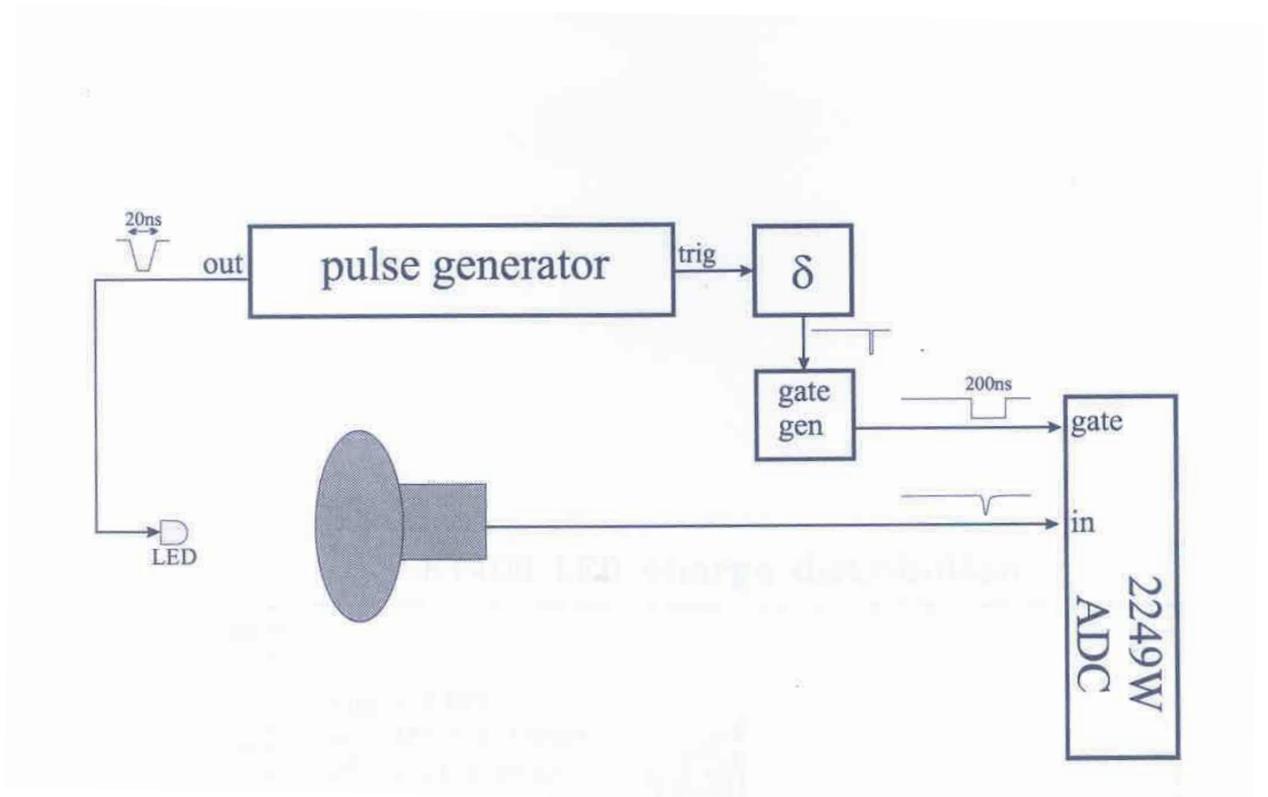}\end{center}
\caption{\label{cap:setup} The test setup scheme.}
\end{figure}

\begin{figure}
\begin{center}\includegraphics[%
 scale=0.5 ]{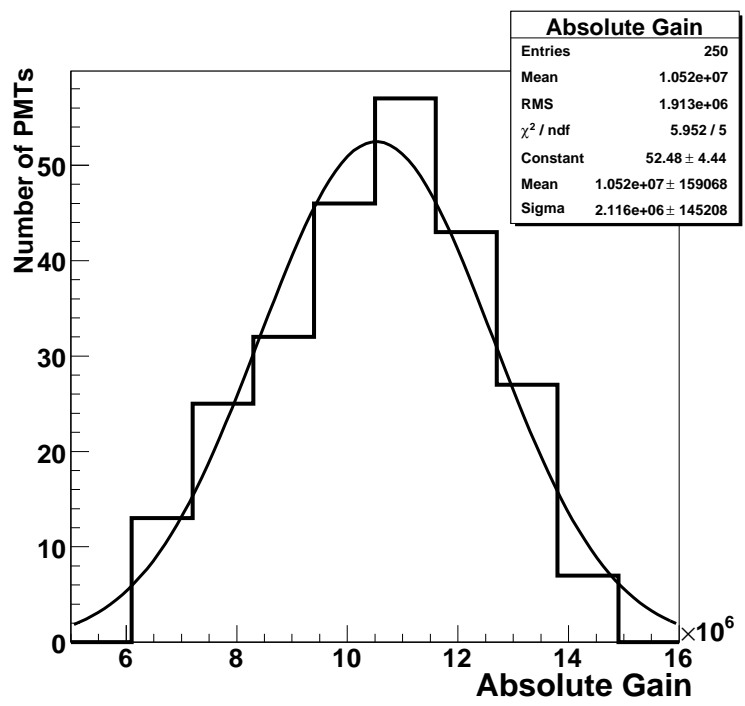}\end{center}
\caption{\label{cap:Gain_OLDBase} The PMTs gain obtained with the
photo-statistics method(using the original linear bases).}
\end{figure}

\begin{figure}
\begin{center}\includegraphics[%
 scale=0.5 ]{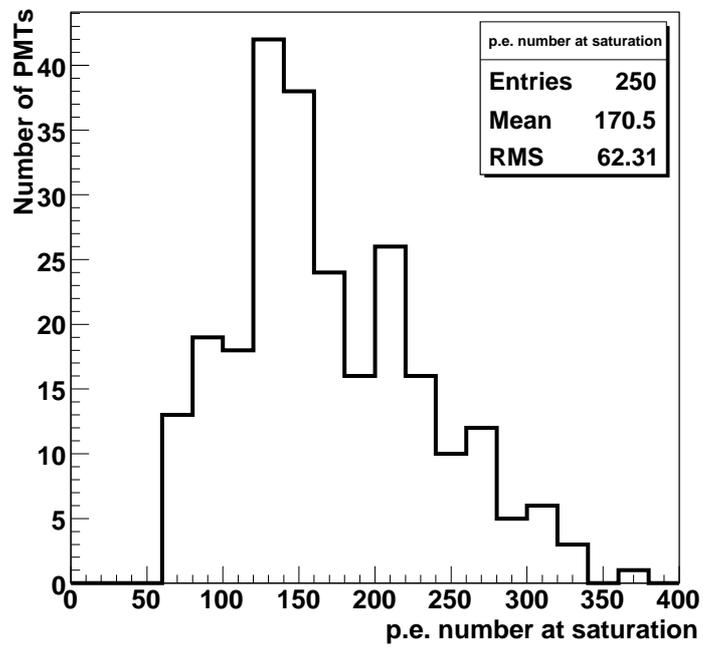}\end{center}
\caption{\label{cap:PESat_OLDBase} The PMTs saturation level obtained with the
photo-statistics method(using the original linear bases).}
\end{figure}

\begin{figure}
\begin{center}\includegraphics[%
 scale=0.6 ]{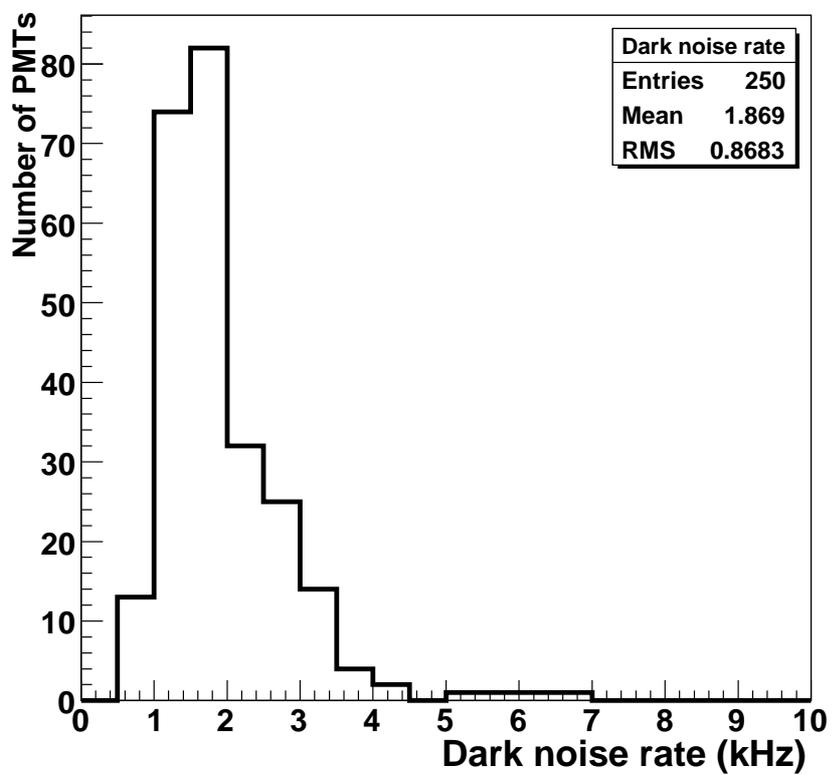}\end{center}
\caption{\label{cap:DR_OLDBase} The dark noise rate (using the original
linear bases).}
\end{figure}

\begin{figure}
\begin{center}\includegraphics[%
 scale=0.6 ]{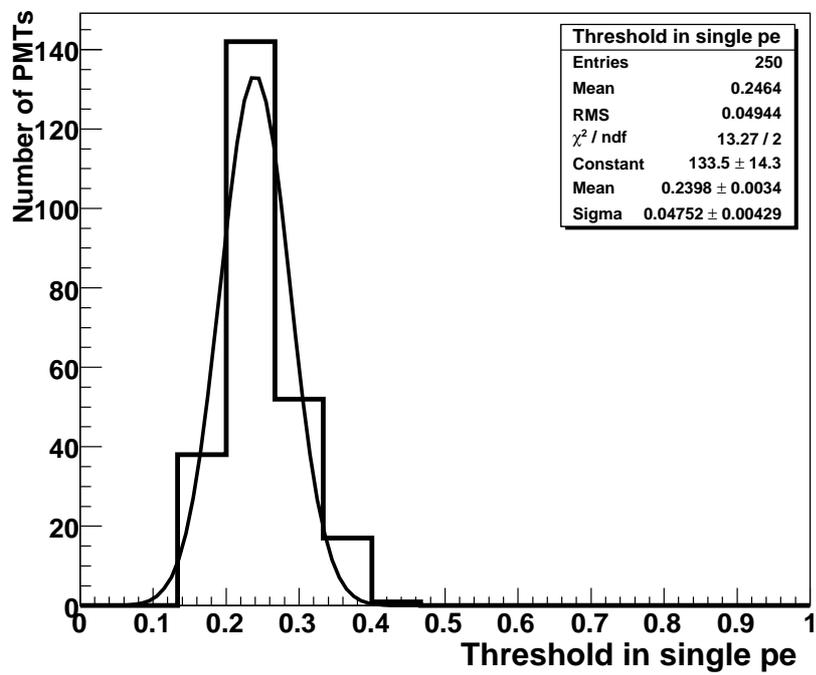}\end{center}
\caption{\label{cap:TrSPE_OLDBase} The threshold in spe at which was measured
the dark noise rate (using the original linear bases).}
\end{figure}

In this section I will present the procedure and the results of the
testing of 250 PMTs, this testing was made with the original old linear bases.
\\
I used the $Am^{241}$ light pulser to determine the high voltage corresponding
to the gain $~10^7$, the result is shown in the figure \ref{cap:HV_OLDBase}.
One time I know the desired high voltage I used the photo-statistics method
to cross check the PMTs gain, to this aim I put each PMT under the
predetermined high voltage, inside the dark box, with an LED connected
to a fast pulser. The trigger output of the pulser was sent to Lecroy 222
Gate and Delay units, first to set a delay, and then to produce a 200 ns
gate that enclosed the PMT pulse that occurred when the LED was flashed,
the used setup is shown in figure \ref{cap:setup}. The PMT signal was
sent to  a Lecroy 2249W ADC, with the gate coming from the 200 ns gate and delay output. The LED was flashed
at 20 Hz at a variety of driving voltages from 2V to 8V, all with very narrow
(20 ns) driving pulse widths. In all cases, the PMT pulse was observed on an
oscilloscope to be completely contained within the 200ns ADC gate.
For each of the ADC distributions, the mean and variance was calculated, as
discussed in section 2, and the ADC variances were plotted as a function of
the the ADC means which let to obtain the gain(figure \ref{cap:Gain_OLDBase})
and the saturation level(figure \ref{cap:PESat_OLDBase}).
To measure the dark noise rate the PMTs were put under high voltage during
12 hours, the signal from the PMT was amplified 100 times and then send to
the input of a LeCroy Model 821 quad discriminator,
the threshold of the discriminator was set
to 2mV, the output of the discriminator was sent to the counter. The
temperature inside the dark box was measured around $19^{\circ} C$.
The measured dark noise rate is shown in the figure \ref{cap:DR_OLDBase}, the
figure \ref{cap:TrSPE_OLDBase} shows the corresponding threshold in single
photo-electron, to evaluate this threshold we need to remember that the dark
noise rate is due to the production of some number, n, of
electrons inside the PMT, without any lighting of the photocathod. These n
produced electrons create a current $I_0$ in the out put of the PMT equal to
$\frac{n \times  e \times G}{t_W}$, where $e=1.6~10^{-19}C,~ t_W$ is
the FWHM of the pulse (measured with a scope), and G is the PMT gain. We can
write $I_0$ as following: $I_0=\frac{V}{Z_0}$, where $V_0$ is the potential(high
of the pulse) and $Z_{0}$ is the impedance of the cable, this lets us to write
n as function of V as following:
\begin{equation}
\label{Threshold-spe}
n=\frac{V \times t_W}{e \times Z_0 \times G}
\end{equation}

The equation \ref{Threshold-spe} link the number of the electrons n
to their potential V, in the
other words this formula lets us to translate the threshold from the volt into
term of number of electrons, in our case V was set to 2mV and $Z_0=50\Omega$.


\newpage

\section{Bases effect on the saturation performance}
As I mentioned at the beginning of this report, these PMTs came with their
own original linear bases and also their own original housing, after that it has been decided to use the tapered bases, the goal is to improve
the saturation performance of the PMTs. There was also a few tests with a new linear bases. In the present section I will discuss the use of this kind of bases.\\

\begin{figure}
\begin{center}\includegraphics[%
 scale=0.6 ]{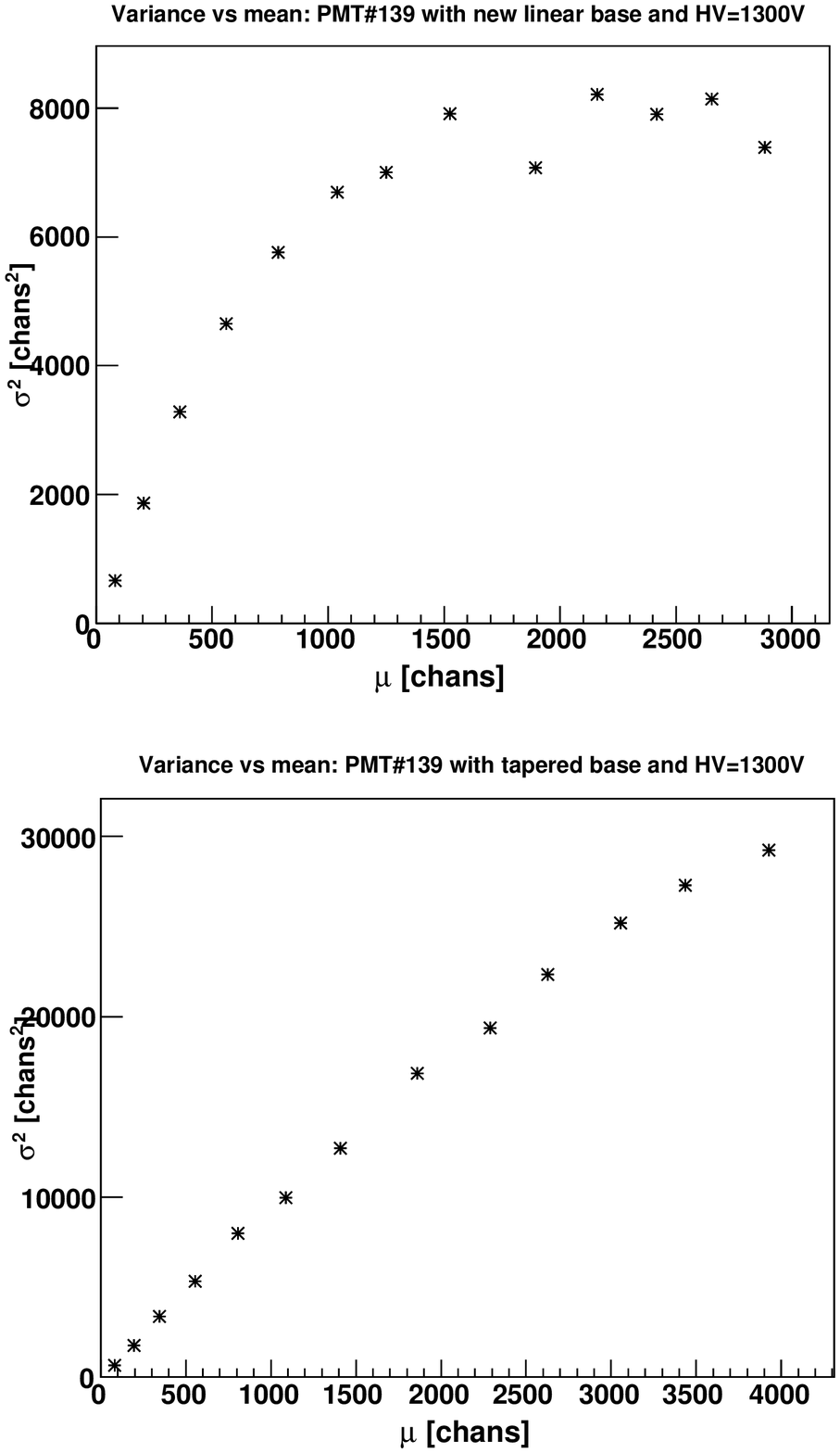}\end{center}
\caption{\label{cap:TapvsLin_pm139} Evolution of the linearity of the PMT
response with the new linear base(on the top), and the new tapered base(on the
bottom).}
\end{figure}
\begin{figure}
\begin{center}\includegraphics[%
 scale=0.6 ]{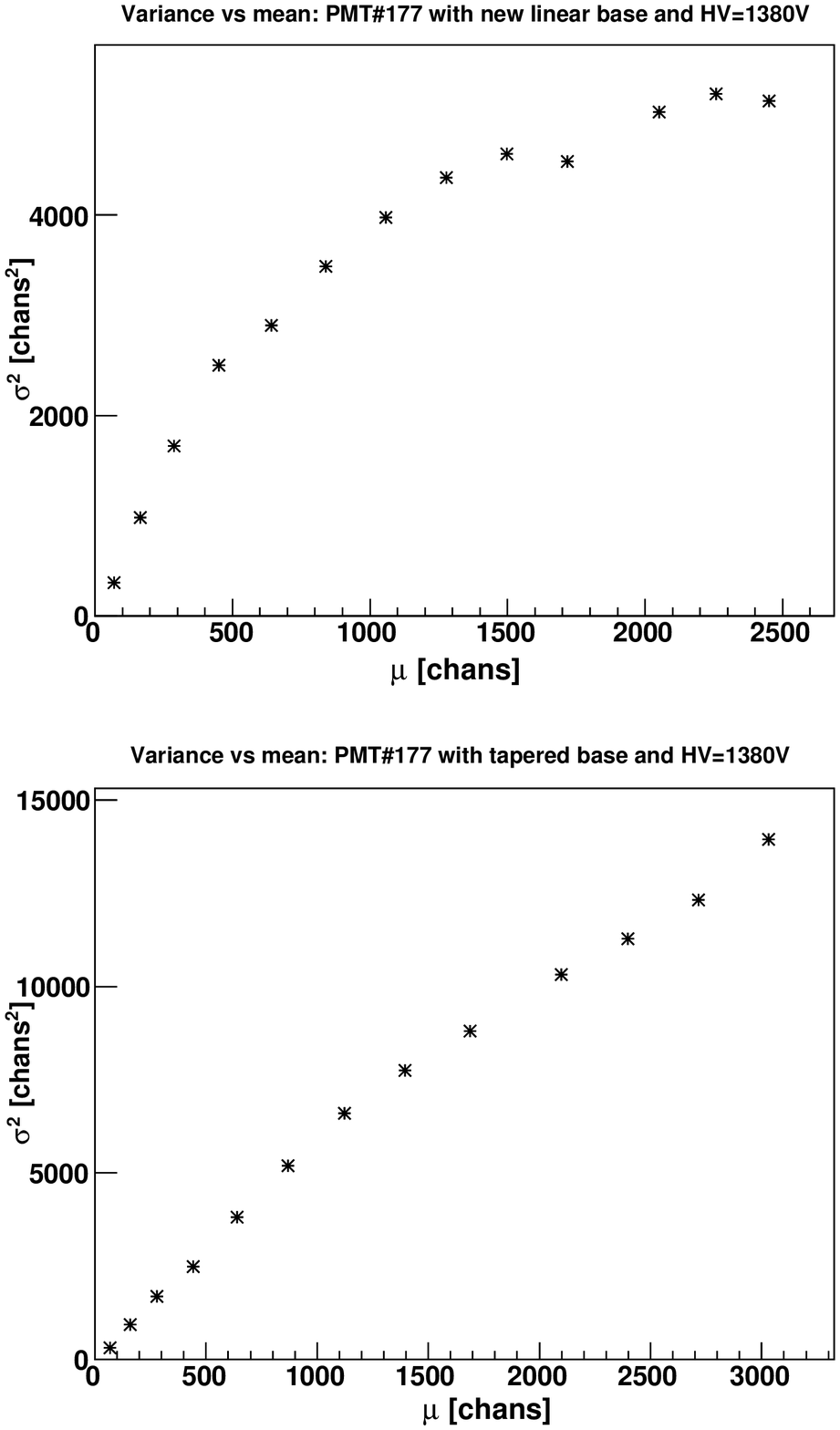}\end{center}
\caption{\label{cap:TapvsLin_pm177} Evolution of the linearity of the PMT
response with the new linear base(on the top), and the new tapered base(on the
bottom).}
\end{figure}
\begin{figure}
\begin{center}\includegraphics[%
 scale=0.5 ]{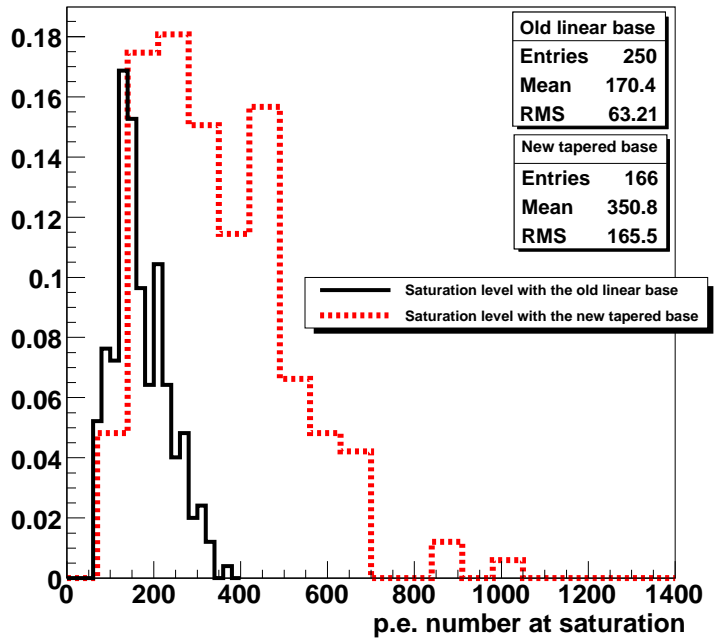}\end{center}
\caption{\label{cap:SatLev-YK2} Comparison of the saturation performance(at the
gain $\sim10^7$) with the old linear bases and the new tapered bases.}
\end{figure}
\begin{figure}
\begin{center}\includegraphics[%
 scale=0.5 ]{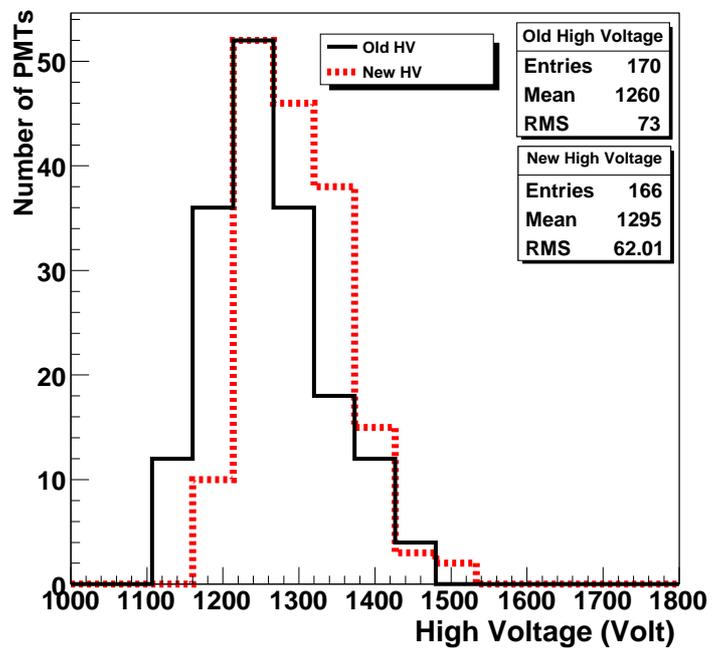}\end{center}
\caption{\label{cap:HVY-K2} Comparison of the high voltage corresponding to the
gain $\sim10^7$, with the old linear bases and the new tapered bases.}
\end{figure}
\begin{figure}
\begin{center}\includegraphics[%
 scale=0.5 ]{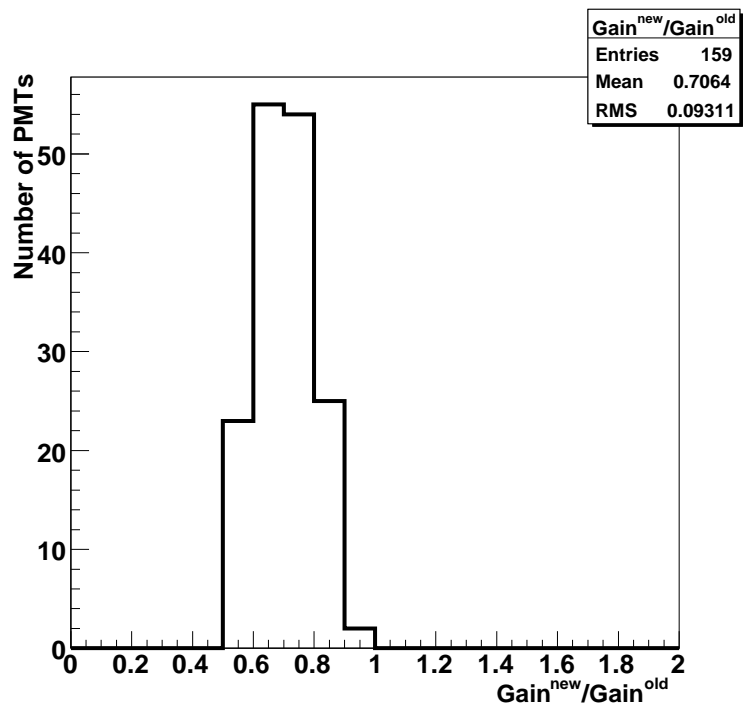}\end{center}
\caption{\label{cap:Gain-ratio} The ratio of the obtained gain with the new
tapered base($Gain^{new}$) and the old linear base($Gain^{old}$).}
\end{figure}
\begin{figure}
\begin{center}\includegraphics[%
 scale=0.5 ]{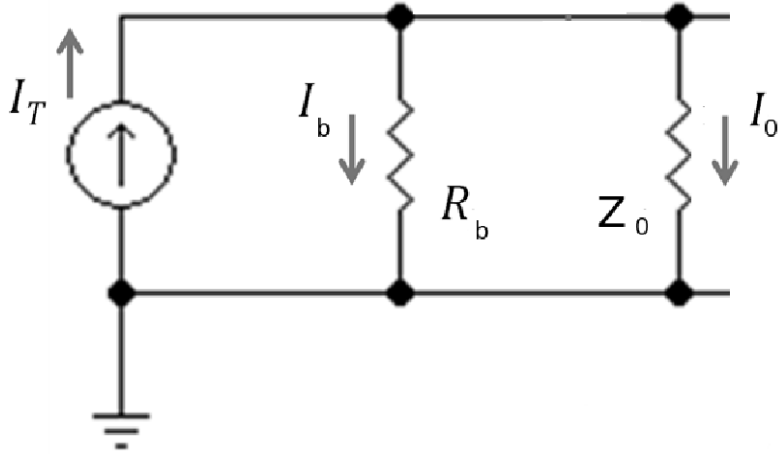}\end{center}
\caption{\label{cap:Current_division} Current divider.}
\end{figure}
\begin{figure}
\begin{center}\includegraphics[%
 scale=0.7 ]{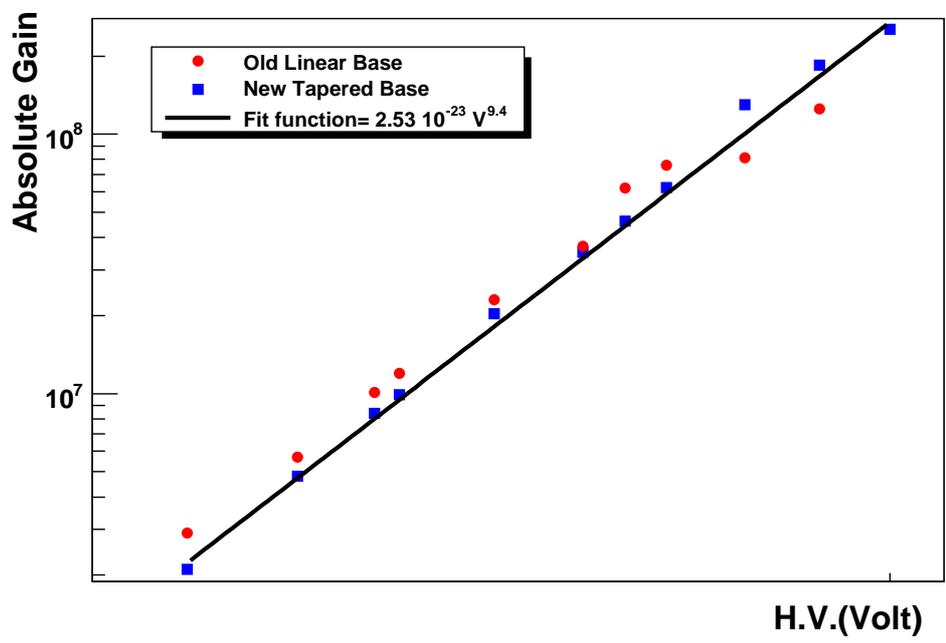}\end{center}
\caption{\label{cap:Gain_HV} The absolute gain, for the PMT 1,  as function as
of the high voltage, the fit is for the data obtained with the tapered base.}
\end{figure}

The figures \ref{cap:TapvsLin_pm139} and
\ref{cap:TapvsLin_pm177} show the plot of the variance square in function of
the ADC mean for two PMTs, with the new linear base and the new tapered base,
its obvious that the region where the PMT response is linear dramatically
extended by a factor 2 or 3. More detailed measurements, for these
two PMTs, are showed in the
table \ref{cap:TapvsLin}, from this table we can make two main
observations: The first is about the gain, with the new bases(linear and
tapered) at the same high
voltage(obtained with the old linear base) the gain decreases, then we need
to increase the high voltage
to reach the desired gain. The second observation is related to the PMT
saturation level which increased by a factor 2 at least. To confirm these
observations I made measurements with more PMTs, the improvement of the
saturation level(measured at the gain $\sim10^7$) is shown in the figure
\ref{cap:SatLev-YK2}. The figure
\ref{cap:HVY-K2} shows the new high voltage corresponding to the gain $\sim10^7$.
The difference in the gain at the same high voltage between the old linear
bases
and the new bases(linear or tapered) is due to the difference in their
back terminations. The old linear bases, as the measurements will show later,
are expected to have a 91$\Omega$
termination instead of 50$\Omega$ used for the new linear and tapered bases,
and all the SHV cables used during this testing have an impedance of
50 $\Omega$. To understand this difference in the gain, we need to remember
that the base termination and the SHV cable work as current divider, like
is illustrated in figure \ref{cap:Current_division}. The current in the
input of the ADC is $I_0$ and it's given by the relation:
\begin{equation}
\label{Idivider}
I_0=\frac{R_b}{Z_0+R_b}I_T
\end{equation}
Where $Z_0$ is the SHV cable impedance, $R_b$ is the back termination of
the base and $I_T$ is the total current. If $I_T$ is constant, it's obvious that
if we change the
base termination and we keep the same cable(same impedance), the current in
the input of the ADC will change and then the gain(measured at the same high
voltage and for the same PMT) will change. In our case the back termination
of the new tapered bases is $R_b=50 \Omega$ and for the old linear bases $R_b$
is expected to be equal to $91 \Omega$ with $5\%$ precision, and
$Z_0=50 \Omega$, then the change in the gain should be:
\begin{equation}
\label{Idiv-gain}
\frac{Gain^{new}}{Gain^{old}}=\frac{R_b^{50}}{R_b^{91}}\frac{Z_0+R_b^{91}}{Z_0+R_b^{50}}=0.775
\end{equation}
To confirm this explanation I show in figure \ref{cap:Gain-ratio} the gain
ratio for the new tapered bases and the old linear bases, the obtained
histogram is wide than expected but its most part is around the
expected value corresponding to $R_b=91 \pm 5\% ~\Omega$, the boundaries of this
histogram can be understood if all the old linear bases don't have the same
back termination, to confirm this hypothesis I measured the back termination
for 11 old linear bases, randomly chosen, the table \ref{cap:Rb-Tap} chows the
 result of these
measurements, as we can see the most part of the old linear bases have the
expected value of the back termination of $91 \Omega$ with respect to the $5\%$
precision, and a few old linear bases have completely different back
termination,
three of them have back termination $\geq 100 \Omega$ and one has $62 \Omega$
back termination. Now, if we take also in account the uncertainty of the
photo-statistics method, used to determine the gain, the histogram of the
figure \ref{cap:Gain-ratio} becomes completely understandable. For more
confirmation and comparisons I took the PMT reference (PMT\# 1) and I put it
under fixed high voltage, after that I measured its gain at the same high
voltage but with different bases (old linear, new linear and new tapered
bases), the table \ref{cap:compare_bases} shows the results of these
these measurements, as I expected I obtained the same gain with the new
linear bases and the new tapered bases because they have the same back
termination, The difference in the back termination lead to difference in the
measured gain not
only between the new bases and the old linear bases, but also between the old
linear bases themselves.\\
I finish this section with short remark on the improvement of the evolution of
the PMT gain as function of the applied high voltage,
the figure \ref{cap:Gain_HV}
shows this evolution, in log-log scale, for the old linear base and
the new tapered base, the high voltage was increased from 1200 V up to 1900 V
for the old linear base and up to 2000 V for the new tapered base,
with a step of
100 V. As we can see the new tapered base gives better linear evolution of
the gain with increased high voltage than the old linear base. Also, the fit of
the measurements with the new tapered base is compatible with 13 stages
amplification. In fact the gain is linked to the applied high voltage in the
following way:$~G=KV^{N\alpha}$, K is a constant which depends on the material
of the dynodes and the voltage division between them, and $\alpha$ is between
0.65 and 0.75. In 13 stages PMT the gain then increases as about the $9^{th}$
power of the applied high voltage, doubling for each 8\% voltage increase.




\begin{sidewaystable}
\begin{tabular}{|c|c|c|c|c|}
\hline
PMT\#&
HV(Volt)&
Abs. Gain($\sim10^7$)&
Number of pe at saturation&
Dark noise rate(kHz)\tabularnewline
\hline
PMT 139(new tapered base)&
1300&
1.13&
508&
1.6 (at 0.221 spe)\tabularnewline
\hline
PMT 139(new linear base)~~~&
1300&
1.04&
189&
1.7(at 0.24 spe)\tabularnewline
\hline
PMT 139(old linear base)~~~&
1220&
1.02&
165&
2.1(at 0.245 spe)\tabularnewline
\hline
\hline
PMT 177(new tapered base)&
1380&
0.74&
468&
1.7(at 0.338 spe)\tabularnewline
\hline
PMT 177(new linear base)~~~&
1380&
0.71&
226&
1.8 (at 0.352 spe)\tabularnewline
\hline
PMT 177(old linear base)~~~&
1380&
1.16&
145&
2.4(at 0.215 spe)\tabularnewline
\hline
\end{tabular}
\caption{\label{cap:TapvsLin} the performances of the new Tapered base
versus the old and the new linear bases.}
\end{sidewaystable}

\begin{table}
\begin{center}\begin{tabular}{|c|c|c|}
\hline
Base\#&
$R^{old}_b (\Omega)$&
$Gain^{new}/Gain^{old}$\tabularnewline
\hline
1&
89&
0.781\tabularnewline
\hline
2&
129&
0.694\tabularnewline
\hline
3&
91&
0.775\tabularnewline
\hline
4&
116&
0.715\tabularnewline
\hline
5&
91&
0.775\tabularnewline
\hline
6&
87&
0.787\tabularnewline
\hline
7&
93&
0.769\tabularnewline
\hline
8&
99&
0.752\tabularnewline
\hline
9&
90&
0.778\tabularnewline
\hline
10&
100&
0.75\tabularnewline
\hline
11&
62&
0.903\tabularnewline
\hline

\end{tabular}\end{center}
\caption{\label{cap:Rb-Tap} The ratio of the obtained gain, with the new
tapered base($Gain^{new}$) and the old linear base($Gain^{old}$), for different
back termination($R^{old}_b$) measured for different old linear bases.}
\end{table}

\begin{sidewaystable}
\begin{tabular}{|c|c|}
\hline
Old linear base\# & Abs. Gain(10\textasciicircum{}7)\tabularnewline
\hline
\hline
1 & 1.37\tabularnewline
\hline
2 & 0.89\tabularnewline
\hline
3 & 1.24\tabularnewline
\hline
4 & 1.09\tabularnewline
\hline
5 & 1.16\tabularnewline
\hline
6 & 1.5\tabularnewline
\hline
\end{tabular}\begin{tabular}{|c|c|}
\hline
New linear base\# & Abs. Gain(10\textasciicircum{}7)\tabularnewline
\hline
\hline
1 & 0.82\tabularnewline
\hline
2 & 0.80\tabularnewline
\hline
\end{tabular}\begin{tabular}{|c|c|}
\hline
New tapered base\# & Abs. Gain(10\textasciicircum{}7)\tabularnewline
\hline
\hline
1 & 0.84\tabularnewline
\hline
2 & 0.84\tabularnewline
\hline
3 & 0.81\tabularnewline
\hline
4 & 0.83\tabularnewline
\hline
5 & 0.84\tabularnewline
\hline
6 & 0.83\tabularnewline
\hline
\end{tabular}

\caption{\label{cap:compare_bases}The measured absolute gain for the same PMT
at the same high voltage, but with different old linear, new linear and tapered
bases.}

\end{sidewaystable}


\newpage

\section{Conclusion}
Different methods were used to make the testing of 250 R1408 PMTs fast and
more accurate, during this testing different type of bases were used, the
new tapered bases gave good saturation performance 2 to 3 times better
than the new linear bases, at the same gain(i.e. the same high voltage). However, at the same
high voltage the PMT gain remains the same with the linear and the tapered bases.
Based on the results of this testing, 156 PMTs were
selected to be used for the inner veto of the Double Chooz detectors.

\section*{Acknowledgments}

I would like to thank Prof. Charles Lane for the interesting discussions  through this work.

\end{document}